\documentclass[preprint]{ptephy_v1}

\preprintnumber{KUNS-2841} 

\usepackage{graphicx}
\usepackage{epsf}
\usepackage{amsmath,amssymb}

\usepackage{float}
\usepackage{color} 
\usepackage{multirow}
\usepackage{dcolumn}
\usepackage{array}
\usepackage{comment}

\usepackage{tikz}
\usepackage{array,booktabs}

\usepackage{ulem}

\def\vector#1{\mbox{\boldmath $#1$}}
\def\hana#1{\mathcal{#1}}
\def\Hesix{{}^6\textrm{He}}
\def\Beten{{}^{10}\textrm{Be}}
\def\Ct{{}^{12}\textrm{C}}
\def\Cf{{}^{14}\textrm{C}}
\def\Os{{}^{16}\textrm{O}}
\def\Oe{{}^{18}\textrm{O}}
\def\2-b{\alpha + \Hesix}
\def\3-b{\alpha + \alpha + 2n}

\def\Kp{K^{\pi}}
\def\GSint{0_\textrm{gs}^+}

\def\NKzero{1_{\textrm{ND}}^-(K=0)}
\def\NKone{1_{\textrm{ND}}^-(K=1)}
\def\CKzero{1_{\textrm{cl}}^-(K=0)}
\def\CKone{1_{\textrm{cl}}^-(K=1)}

\usepackage[utf8]{inputenc}

\usepackage{natbib}

\begin{document}
\title{Low-energy dipole excitation mode in $^{18}$O with antisymmetrized molecular dynamics
}

\author{Yuki Shikata}
\author{Yoshiko Kanada-En'yo}
\affil{Department of Physics, Kyoto University, Kyoto 606-8502, Japan}


\begin{abstract}
Low-energy dipole (LED) excitations in $\Oe$ were investigated using a combination of the variation after $K$-projection in the framework of antisymmetrized molecular dynamics with $\beta$-constraint with the generator coordinate method.
We obtained two LED states, namely, the $1_1^-$ state with a dominant shell-model structure and the $1_2^-$ state with a large $\Cf+\alpha$ cluster component. 
Both these states had significant toroidal dipole (TD) and compressive dipole (CD) strengths,
indicating that the TD and CD modes are not separated but mixed in the LED excitations of $\Oe$.
This is unlike the CD and TD modes for well-deformed nuclei such as  $^{10}$Be, where the CD and TD modes are generated as $\Kp=0^-$ and $\Kp=1^-$ excitations, respectively, in a largely deformed state.
\end{abstract}

\subjectindex{D11, D13}

\maketitle
\section{Introduction}
 
Low-energy dipole (LED) excitation has gained considerable interest among both experimental and theoretical researchers for few decades~\cite{Paar:2007bk,1402-4896-2013-T152-014012,Bracco:2015hca}.
LEDs have been observed in a lower energy region than giant dipole resonances and have significant dipole strengths of about several percent of the energy-weighted sum rule (EWSR) 
in $N=Z$ nuclei such as $^{12}$C~\cite{John:2003ke} and $^{16}$O~\cite{Harakeh:1981zz}.
LEDs were recently discovered in neutron-rich nuclei such as $^{20}$O~\cite{Tryggestad:2002mxt,Tryggestad:2003gz,Nakatsuka:2017dhs}, $^{26}$Ne~\cite{Gibelin:2008zz}, and $^{48}$Ca~\cite{PhysRevLett.85.274,Derya:2014yqk}.
Although LED modes were investigated in many theoretical studies, 
the observed LED strengths were not fully described, and LED properties are still not well understood.
Kvasil {\it et al}. introduced the toroidal dipole (TD) and compressive dipole (CD) operators to prove the vortical and compressional modes, respectively~\cite{0954-3899-29-4-312}, and predicted 
the existence of the vortical~(toroidal) mode in some neutron-rich nuclei~\cite{Kvasil:2011yk,Repko:2012rj,Reinhard:2013xqa}.
Chiba {\it et al}. showed that the cluster excitation can enhance CD strength and contribute to the LED mode~\cite{Chiba:2015khu}.

LED excitations in oxygen isotopes have attracted theoretical and experimental researchers for a few decades.
For $^{17-22}$O, the isovector dipole strengths were observed in the low-energy region;
a few percent of Thomas–Reiche–Kuhn sum rule were found below the excitation energy of 15\ MeV~\cite{Leistenschneider:2001zz}.
Recently,  Nakatsuka {\it et al}. measured the significant isoscalar (IS) LED strengths 
in $^{20}$O~\cite{Nakatsuka:2017dhs}.
To understand the properties of LED excitations in neutron-rich oxygen isotopes, 
theoretical studies were conducted based on the mean-field approach, and 
LEDs were described as nonresonant single-particle excitations of weakly bound neutrons~\cite{Sagawa:2001jhf, Vretenar:2001hs, Paar:2002gz}.
On the other hand, origins besides single-particle excitations can be considered for the excited states of 
O isotopes. 
For example, cluster structures have been intensively discussed for $\Oe$ in experimental and theoretical studies. 
Gai {\it et al.} observed $E1$, $E2$, and $\alpha$-widths for low-energy states and proposed cluster bands,  including $1^-$~(4.46\ MeV) and $1^-$~(6.20\ MeV) states, with a large $\Cf+\alpha$ cluster component~\cite{Gai:1983zz, Gai:1987zz, PhysRevC.43.2127}.
In experiments on $\Cf+\alpha$ elastic scattering, many excited states with large $\alpha$-spectroscopic factors were observed in the energy region below 14.9\ MeV~\cite{Johnson;2009j,Avila:2014zwa}. 
Further in theoretical studies, cluster states, including $\Cf+\alpha$ cluster and $\Ct+\alpha+2n$ molecular structures in $\Oe$ were examined using cluster models~\cite{Baye:1984ljb, Descouvemont:1985zz} and antisymmetrized molecular dynamics (AMD)~\cite{Furutachi:2007vz,Baba:2019csd,Baba:2020iln}.
However, these cluster structures have not been discussed in association with LED excitations.

In recent years, LED excitations in deformed systems were theoretically examined\cite{Nesterenko:2017rcc, Nesterenko:2016qiw, Kvasil:2013yca,Chiba:2019dap, Shikata:2020lgo}. 
Nesterenko {\it et al}. discussed the coexistence of the TD and CD modes in the lower energy region of deformed nuclei such as $^{24}$Mg~\cite{Nesterenko:2017rcc}, $^{132}$Sn~\cite{Nesterenko:2016qiw}, and $^{170}$Yb~\cite{Kvasil:2013yca}. 
In our previous work~\cite{Shikata:2020lgo}, we extended the AMD method for studying LED and 
applied the method with variation after $K$-projection (K-VAP) to $\Beten$ and $\Os$.
The AMD with K-VAP was useful for studying two types of LED, namely, the TD and CD modes
in deformed systems by treating separately the $K=1$ and $K=0$ components and their mixing. 
(The $K$-quantum number is defined by the $Z$-component of the total angular momentum in the body-fixed frame.)
For $\Beten$, we found that significant TD strength was generated by the $K=1$ component, 
showing a remarkable vortical nature. 
An interesting result is that TD and CD modes are clearly separated by the $K$-quantum number in the $1_1^-$ and $1_2^-$ states of $\Beten$ because of large deformation with a developed cluster structure.
However, this is not so in the case of $\Os$, where two modes are mixed in the $1_1^-$ and $1_2^-$ states.

In this study, we investigated LED excitations in $\Oe$ by applying the same method of AMD as used in the previous paper~\cite{Shikata:2020lgo}.
Namely, we applied the $\beta$-constrained AMD with K-VAP, which was combined with the generator coordinate method (GCM). 
We aimed to clarify the role of cluster structures and determine whether the TD and CD modes appear
in LED states in $\Oe$. 
The TD and CD strengths in the LED states are analyzed, and the vortical nature was discussed. 

This paper is organized as follows.
In Sect.~\ref{sec:formalism}, the framework of $\beta$-AMD with K-VAP and GCM are explained.
The definition of the dipole operators are also given. 
Section~\ref{sec:effective_int} describes the effective nuclear interaction used in the present calculation.
The calculated results for $\Oe$ are shown in Sect.~\ref{sec:results}.
The properties of LED excitations in $\Oe$ are discussed in Sect.~\ref{sec:discussion}.
Finally, a summary is given in Sect.~\ref{sec:summary}.

\section{Formalism}\label{sec:formalism}

To investigate LED excitation, we apply K-VAP in the framework of $\beta$-AMD to $\Oe$.
Total wave functions of $\Oe$ are obtained by GCM calculation for $\beta$ values and for $K$-mixing.
We calculate the dipole transition strengths for three dipole operators to $1^-$ states.

\subsection{$\beta$-constraint AMD with K-VAP}

An AMD wave function for $A$-body system $\Phi$ is expressed by a Slater determinant of single particle wave functions~\cite{Kanada-Enyo:1998onp,Kanada-Enyo:2001yji}:
\begin{eqnarray}
\Phi = \hana{A}\left[\psi_1\psi_2\cdots\psi_A\right]
\end{eqnarray}
where $\psi_i$ represents the $i$th single particle wave function written as follows:
\begin{eqnarray}
\psi_i\ &=& \phi(\vector{Z}_i)\chi(\vector{\xi}_i)\tau_i, \\
\phi(\vector{Z}_i) &=& \left(\frac{2\nu}{\pi}\right)^{\frac{3}{4}}\exp\left[ -\nu\left(\vector{r} - \frac{\vector{Z}_i}{\sqrt{\nu}}\right)^2 \right],\\
\chi(\vector{\xi}_i)&=&\xi_{i\uparrow}|\uparrow\rangle + \xi_{i\downarrow}|\downarrow\rangle, \\
\tau_i &=&p\ \textrm{or}\ n.
\end{eqnarray}
The spatial part of the single particle wave function is given by a localized Gaussian wave packet.
Here $\vector{Z_i}$ and $\vector{\xi_i}$ are the parameters of the Gaussian centroids and spin directions, respectively, and are treated as complex variational parameters.
The width parameter $\nu$ is common for all nucleons.

In the AMD, to obtain the base function, the energy variation 
\begin{eqnarray}
\delta\left(\frac{\langle\Psi|\hat{H}|\Psi\rangle}{\langle\Psi|\Psi\rangle} \right) = 0 \label{eq:var}
\end{eqnarray}
is performed for the effective hamiltonian $\hat{H}$.
In the K-VAP formalism proposed in our previous paper~\cite{Shikata:2020lgo}, the energy variation is done for the parity- and $K$-projected AMD wave function $|\Psi\rangle = \hat{P}_K\hat{P}^{\pi}|\Phi\rangle$, where $\hat{P}^{\pi}$ is the parity-projection operator, and $\hat{P}_{K}$ is the $K$-projection operator given as
\begin{eqnarray}
\hat{P}_K = \frac{1}{2\pi}\int_0^{2\pi}d\theta\ e^{-iK\theta}\hat{R}(\theta),
\end{eqnarray}
where $\hat{R}(\theta)$ is the rotation operator around the principal-axis in body-fixed frame.
With the K-VAP method, the wave function optimized for each $K^{\pi}$ can be obtained.
In the present work, to obtain the basis wave functions for the ground and dipole states in $\Oe$, we perform K-VAP with the $\Kp=0^+,\ \Kp=0^-$, and $\Kp=1^-$ projections and obtain the bases called $\Kp=0^+,\ \Kp=0^-$, and $\Kp=1^-$ bases, respectively.

To consider the quadrupole deformations of an AMD wave function, 
we use the deformation parameters $\beta$ and $\gamma$ defined as follows:
\begin{eqnarray}
\beta\cos\gamma &=& \frac{\sqrt{5}}{3}\frac{2\langle z^2\rangle - \langle x^2\rangle - \langle y^2\rangle}{R^2}, \\
\beta\sin\gamma &=& \sqrt{\frac{5}{3}}\frac{\langle x^2\rangle - \langle y^2\rangle}{R^2}, \\
R^2 &=& \frac{5}{3}(\langle x^2\rangle + \langle y^2\rangle + \langle z^2\rangle),
\end{eqnarray}
where $\langle r_\sigma^2\rangle$ $(\sigma=x,y,z)$ is the expected value of the one-body operator $\hat{r}_\sigma^2=\frac{1}{A}\sum_{i=1}^A\hat{r}_{i\sigma}^2$ for the AMD wave function before the projections.
In the framework of $\beta$-constraint AMD, 
we perform the $K$-projected energy variation under the $\beta$-constraint 
but no constraint on $\gamma$  
to obtain the AMD wave function $|\Phi_K^\pi(\beta)\rangle$ optimized for a given $\beta$ value~\cite{Dote:1997zz,10.1143/PTP.106.1153}.
Thus, sets of basis AMD wave functions $|\Phi_{K=0}^+(\beta)\rangle$, $|\Phi_{K=0}^-(\beta)\rangle$, 
$|\Phi_{K=1}^-(\beta)\rangle$ for various $\beta$ values are obtained.

\subsection{GCM}
To obtain the total wave function of the $J^{\pi}_m$ state of $\Oe$, 
the obtained wave functions are superposed by $K$-mixing and GCM with respect to the generator coordinate $\beta$ as follows:
\begin{eqnarray}
\Psi^{\pi}(J_m) = \sum_{K,K'}\sum_{\beta}c_{KK'}(\beta)\hat{P}_{MK'}^J\hat{P}^{\pi}|\Phi_K^{\pi}(\beta)\rangle ,
\end{eqnarray}
where $\hat{P}_{MK}^J$ is the angular-momentum-projection operator.
Coefficients $c_{KK'}(\beta)$ are determined by diagonalizing the Hamiltonian and norm matrices.
Note that for the negative parity states, $\Kp=0^-$ and $\Kp=1^-$ bases are also superposed in addition to $K$-mixing.

\subsection{Dipole operators}
We calculate the transition strengths of three dipole operators, namely, $E1$, TD, and CD operators, for transitions from the ground state to the LED states:
\begin{eqnarray}
\hat{M}_{E1}(\mu) &=& \frac{N}{A}\sum_{i\in p}r_iY_{1\mu}(\hat{\vector{r}}_{i}) - \frac{Z}{A}\sum_{i\in n}r_iY_{1\mu}(\hat{\vector{r}}_{i}), \\
\hat{M}_{\textrm{TD}}(\mu)&=&\frac{-1}{10\sqrt{2}c} \int d\vector{r}\ (\nabla\times\vector{j}_{\textrm{nucl}}(\vector{r}))\cdot r^3\vector{Y}_{11\mu}(\hat{\vector{r}}), \label{eq:TD_op}\\
\hat{M}_{\textrm{CD}}(\mu) &=& \frac{-1}{10\sqrt{2}c}\int d\vector{r}\ \nabla\cdot\vector{j}_{\textrm{nucl}}(\vector{r})\ r^3Y_{1\mu}(\hat{\vector{r}}), \label{eq:CD_op}
\end{eqnarray}
where $\vector{Y}_{jL\mu}(\hat{\vector{r}})$ is vector spherical~\cite{0954-3899-29-4-312}.
Further, $\vector{j}_{\textrm{nucl}}(\vector{r})$ is the convection nuclear current defined by
\begin{eqnarray}
\vector{j}_{\textrm{nucl}}(\vector{r}) &=& \frac{-i\hbar}{2m}\sum_{k=1}^A\{ \vector{\nabla}_k\delta(\vector{r}-\vector{r}_k) + \delta(\vector{r}-\vector{r}_k)\vector{\nabla}_k \}.  \label{eq:current}
\end{eqnarray}
$E1$ operator measures the isovector dipole mode, whereas 
the TD and CD operators measure the nuclear vorticity~\cite{Kvasil:2011yk} and nuclear compressional mode, respectively.

The transition strength of a dipole operator $D=\{E1, \textrm{TD},\textrm{CD}\}$ 
for $0^+_1\to 1^-_k$ is given as
\begin{eqnarray}
B(D;0_1^+\rightarrow 1_k^-) = |\langle 1_k^-||\hat{M}_D||0_1^+\rangle|^2 .
\end{eqnarray}
It is noted that the CD transition strength is consistent with the standard ISD transition strength:
\begin{eqnarray}
B(\textrm{CD};0_1^+\rightarrow 1_k^-) &=& \left(\frac{1}{10}\frac{E_k}{\hbar c}\right)^2B(\textrm{ISD};0_1^+\rightarrow 1_k^-),
\end{eqnarray}
where $E_k$ is the excitation energy of the $1_k^-$ state.

\section{Effective interaction} \label{sec:effective_int}
The effective Hamiltonian used in the present study is given as
\begin{eqnarray}
H = \sum_i t_i - T_G + \sum_{i<j}v_{ij}^{\textrm{coulomb}} + V_{\textrm{eff}}.
\end{eqnarray} 
Here, $t_i$ and $T_G$ are the kinetic energy of the $i$th nucleon and that of the center of mass, respectively,
 and $v_{ij}^{\textrm{coulomb}}$ is the Coulomb potential.
The effective nuclear potential $V_{\textrm{eff}}$ includes the central and spin-orbit potentials.
We use the MV1(case 1) central force~\cite{Ando:1980hp} with the parameters $W = 1-M = 0.62$ and $B=H=0$, and 
the spin-orbit part of the G3RS force~\cite{Tamagaki:1968zz, Yamaguchi:1979hf} with the strengths $u_1=-u_2=-3000$\ MeV.
This set of parametrization is identical to that used for the AMD calculations of $p$-shell and $sd$-shell nuclei
in Refs.~\cite{Kanada-Enyo:1999bsw,Kanada-Enyo:2006rjf,Kanada-Enyo:2017ers,Kanada-Enyo:2020goh}.
It describes the energy spectra of $\Ct$ including the $1^-$ states.
The width parameter is chosen as $\nu=0.16$ fm$^{-2}$, which reproduces the nuclear size of $^{16}$O 
with a closed $p$-shell configuration.

\section{Results} \label{sec:results}

In this section, we show the results of $^{18}$O calculated by $\beta$-AMD with K-VAP.
We mainly discuss the structure of the $0^+$ and $1^-$ states.

\subsection{Energy curves and intrinsic structure}

\begin{figure}[!h]
\begin{center}
\includegraphics[width=12cm]{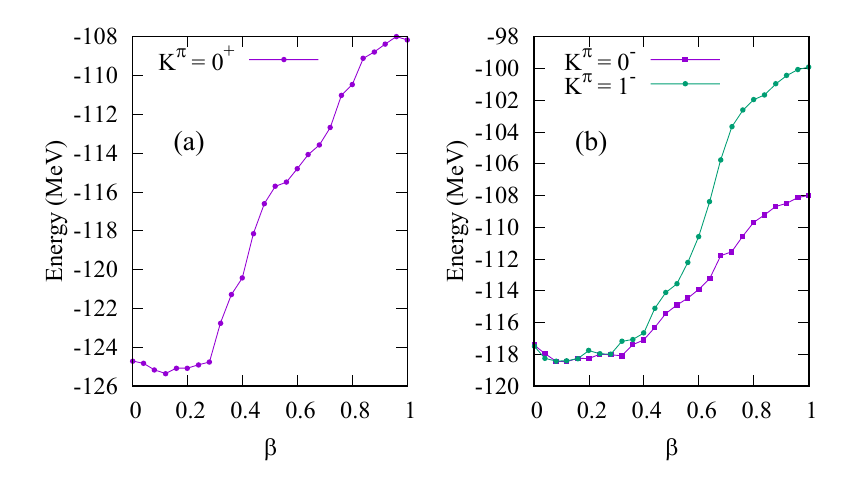}
\end{center}
\caption{$K$-projected energy curves as a function of $\beta$ in $\Oe$ obtained by $\beta$-AMD with K-VAP.
(a) The energy of the $\Kp=0^+$ base with $\Kp=0^+$ projection.
(b) The energy of the $\Kp=0^-$ and $\Kp=1^-$ bases with $\Kp=0^-$ and $\Kp=1^-$ projections, respectively.
}
\label{fig:energy_surface}
\end{figure}

By applying the $\beta$-AMD method with K-VAP, we obtain the energy minimum bases 
for $K^\pi=0^+$, $K^\pi=0^-$, and $K^\pi=1^-$ at given $\beta$ values.
The $K$-projected energy curve obtained for $\Kp=0^+$ is shown in Fig.~\ref{fig:energy_surface}~(a), and the 
 $\Kp=0^-$ and $\Kp=1^-$ energy curves are shown in Fig~\ref{fig:energy_surface}~(b).
In all cases of $\Kp=0^+$, $\Kp=0^-$, and $\Kp=1^-$, the $\Kp$ energy curve has the minimum in the small $\beta$ region ($\beta<0.4$) and no local minimum in the large $\beta$ region ($\beta>0.4$).
For the negative parity, the $\Kp=0^-$ and $\Kp=1^-$ energies are almost consistent with each other 
in the small $\beta$ region.
In this region, the $K$-quantum number is not well defined and 
the $\Kp=0^-$ and $\Kp=1^-$ projected states are similar to each other. 
In the large $\beta$ region ($\beta>0.4$), the $\Kp=0^-$ and $\Kp=1^-$ energies are divided.
The $\Kp=0^-$ energy is lower than the  $\Kp=1^-$ energy, implying that the $\Kp=0^-$ component is favored in this $\beta$ region. 
We emphasize that the $\Kp=1^-$ component in the large $\beta$ region contains excited configurations optimized for quanta $\Kp=1^-$, which cannot be obtained by the usual $\beta$-constraint variation without the $\Kp$ projection. This is an advantage of the present K-VAP method. 

\begin{figure}[!h]
\begin{center}
\includegraphics[width=\hsize]{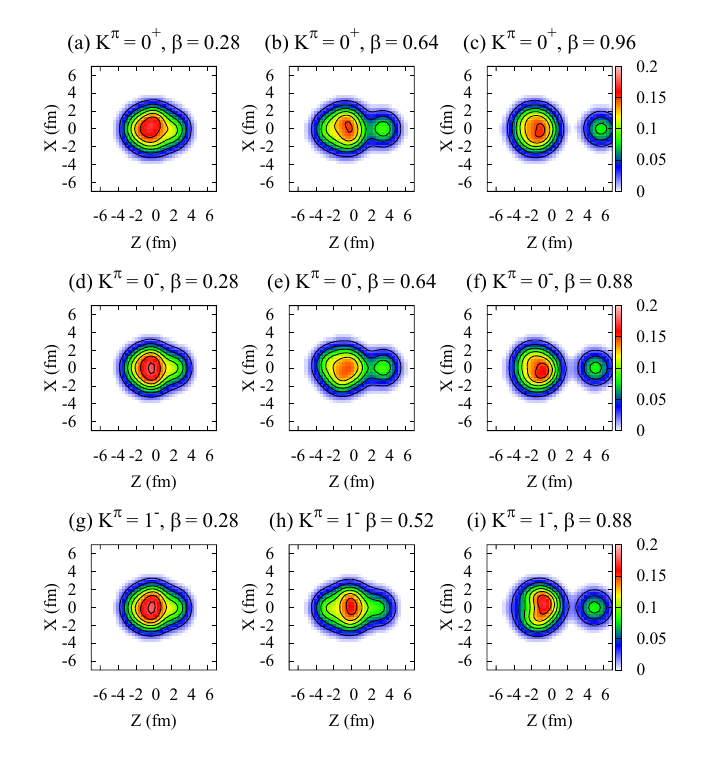}
\end{center}
\caption{Intrinsic density distributions of the basis wave functions, 
(a)~$\Phi_{K=0}^+(\beta=0.28)$, (b)~$\Phi_{K=0}^+(\beta=0.64)$, (c)~$\Phi_{K=0}^+(\beta=0.96)$, 
(d)~$\Phi_{K=0}^-(\beta=0.28)$, (e)~$\Phi_{K=0}^-(\beta=0.64)$, (f)~$\Phi_{K=0}^-(\beta=0.88)$,
(g)~$\Phi_{K=1}^-(\beta=0.28)$, (h)~$\Phi_{K=1}^-(\beta=0.52)$, and (i)~$\Phi_{K=1}^-(\beta=0.88)$,
obtained by K-VAP with the $\beta$-constraint. 
The density is projected onto the $Z-X$ plane by integrating along the $Y$-axis.
}
\label{fig:density}
\end{figure}

For each $\beta$, three configurations are obtained by the K-VAP calculation 
for the positive~($\Kp=0^+$) and negative~($\Kp=0^-$ and $\Kp=1^-$) parities.
The intrinsic density distributions of these bases are shown in Fig.~\ref{fig:density}.
For positive parity, the $\Kp=0^+$ bases at $\beta=0.28$, $\beta=0.64$, and $\beta=0.96$ are shown in Figs.~\ref{fig:density}~(a), (b), and (c), respectively.
The shell model states are obtained for the small $\beta$ region corresponding to the energy minimum (Fig.~\ref{fig:density}~(a)).
This configuration dominantly contributes to the ground state.
The $\Cf+\alpha$ cluster bases obtained in the deformed region of $\beta\sim0.6$ contribute to
the $0_2^+$ band, which is regarded as the first $\Cf+\alpha$ cluster band.
With increase in the deformation, the clustering around $\beta\sim 1.0$ developed further, as seen in Fig.~\ref{fig:density}~(c), and constructs a higher-nodal $\Cf+\alpha$ cluster band of the 
$0_4^+$ state.
For the negative parity, the density distributions of the $\Kp=0^-$ bases at $\beta=0.28$, $\beta=0.64$,
and $\beta=0.88$ are shown in Figs.~\ref{fig:density}~(d), (e), and (f), respectively, 
and those of the $\Kp=1^-$ bases at $\beta=0.28$, $\beta=0.52$,
and $\beta=0.88$ are shown in Figs.~\ref{fig:density}~(g), (h), and (i), respectively. 
The shell model states obtained for the small $\beta$ ($\leq0.4$) region are shown in Figs.~\ref{fig:density}~(d) and (g). 
The $\Kp=0^-$ and $\Kp=1^-$ components of these bases are almost identical to each other and 
give dominant contributions to the $1_1^-$ state.
For the $K^\pi=0^-$ bases in the $\beta\gtrsim 0.4$ region, 
the developed $\Cf+\alpha$ cluster bases are obtained (see Figs.~\ref{fig:density}~(e) and (f));
they contribute to the $1_2^-$ and $1_3^-$ states, 
which can be regarded as the parity partners of the $0_2^+$ and $0_4^+$ states in the cluster bands, respectively. 
For the $\Kp=1^-$ bases at $\beta\sim 0.5$, an $\alpha$ cluster is formed at the surface of the $\Cf$ cluster
but the reflection asymmetry in the $Z$ direction is not as remarkable as that of the $\Kp=0^-$ bases.
This finding indicates that these $\Kp=1^-$ components mainly contain single-particle excitation instead of the parity asymmetric $\Cf+\alpha$-cluster excitation. 
In the $\beta\gtrsim 0.6$ region, the $\Kp=1^-$ bases show $\Cf+\alpha$-like structures, but the $\Cf$ cluster is somewhat distorted from the almost spherical $\Cf$ clusters in the $\Kp=0^+$ and $\Kp=0^-$ bases.

\subsection{GCM results: Energy spectra and LED strengths}

\begin{figure}[!h]
\begin{center}
\includegraphics[width=\hsize]{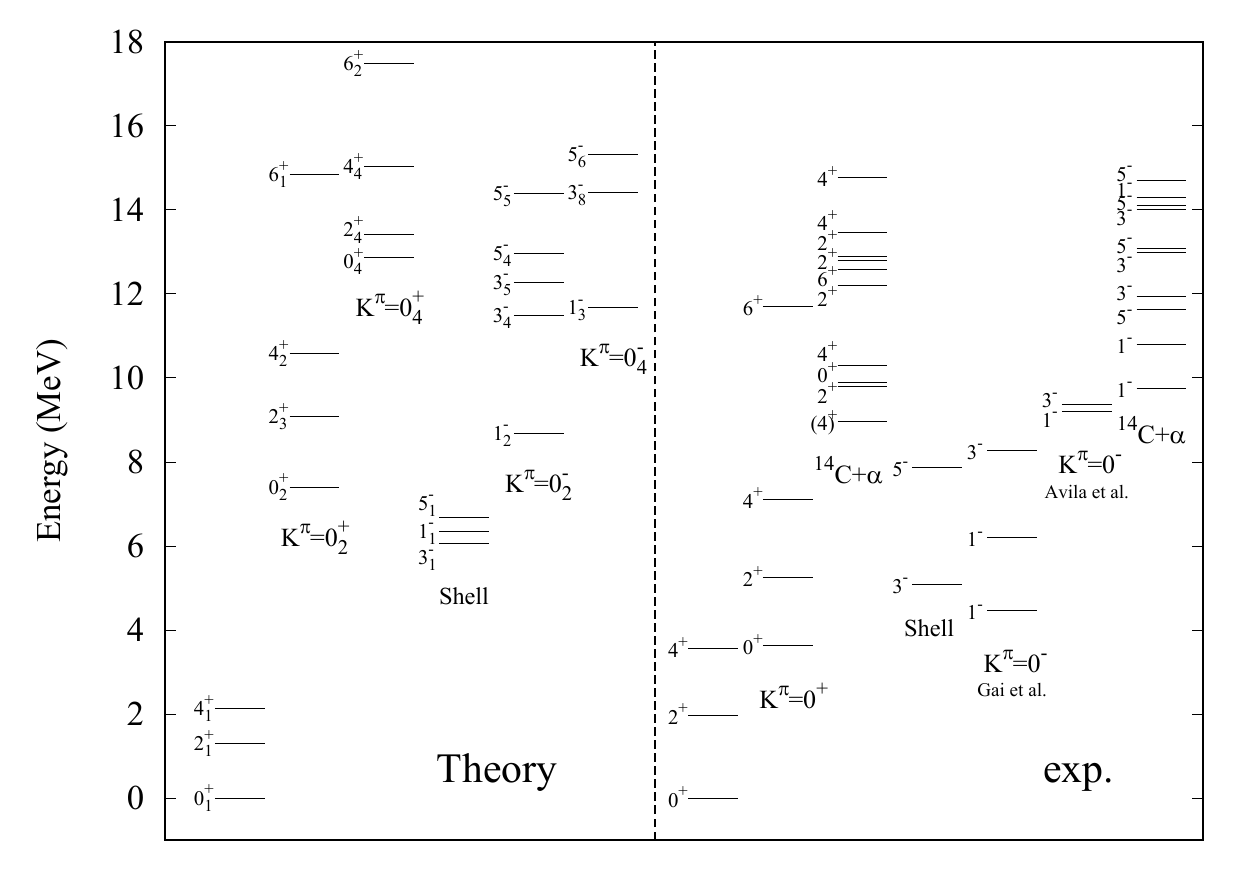}
\end{center}
\caption{
Calculated and experimental energy spectra of $\Oe$.
For the calculated spectra, the ground-band states and excited states assigned to 
the $\Cf+\alpha$ and $\Cf+\alpha$ (higher nodal) bands labeled $\Kp=0_2^{\pm}$ and $\Kp=0_4^{\pm}$, respectively, are shown together with the low-lying shell model states, i.e., $1_1^-$, $3_1^-$, and $5_1^-$ states.
For the experimental spectra, the ground band states and cluster-band candidates 
for $\Kp=0^+$~\cite{Cunsolo:1981zza,Avila:2014zwa} and $\Kp=0^-$ bands proposed by Gai {\it et al}.~\cite{Gai:1983zz,PhysRevC.43.2127} and Avila {\it et al}.~\cite{Avila:2014zwa} are shown.
The states labeled $\Cf+\alpha$ are those with $\alpha$-spectroscopic factor $\theta_{\alpha}^2\geq 0.09$
in $E_x\leq 14.9$~MeV as reported in Ref.~\cite{Avila:2014zwa}. 
The low-lying shell model states, i.e., the $3^-$(5.10\ MeV) and 
$5^-$(7.96\ MeV) states, are also shown in the experimental spectra.
}
\label{fig:spectrum}
\end{figure}

From the GCM calculation, the binding energy for the ground state is obtained as $128.8$\ MeV, which is underestimated compared to the experimental value of $139.8$\ MeV.
The energy spectra of $\Oe$ obtained by the GCM calculation are shown in Fig.~\ref{fig:spectrum}.
In addition to the calculated spectra, the observed spectra up to 14.9\ MeV are also shown.
For the calculated spectra, we show the ground-band states and excited states assigned to 
the lowest and higher nodal $\Cf+\alpha$-cluster bands labeled $\Kp=0_2^{\pm}$ and $\Kp=0_4^{\pm}$, respectively. 
The low-lying shell model states, the $1_1^-$, $3_1^-$, and $5_1^-$ states are also shown.
In the experimental spectra, we show the experimental states labeled "$\Kp=0^+$" and "$\Kp=0^-$" are cluster-band candidates for the $\Kp=0^+$ and $\Kp=0^-$ bands, respectively~\cite{Cunsolo:1981zza,Gai:1983zz,PhysRevC.43.2127,Avila:2014zwa}.
We also show the experimental spectra for the ground band and low-lying shell model states, i.e., the $3^-$(5.10\ MeV) and $5^-$(7.96\ MeV) states, and those for other $\Cf+\alpha$-cluster states 
with the $\alpha$-spectroscopic factor $\theta_{\alpha}^2\geq 0.09$ reported in Ref.~\cite{Avila:2014zwa}.

The ground band is dominated by the shell model bases obtained for $\Kp=0^+$ 
at small $\beta$; the $0^+_1$ state has 93~\% overlap with the $\Kp=0^+$ base at $\beta=0.28$.
We obtain the $\Cf+\alpha$-cluster band starting from the $0_2^+$ state at $7.40$\ MeV, which is mainly 
constructed by the $\Kp=0^+$ bases in the $\beta=0.5$--$0.7$ region.   
Furthermore, the higher-nodal $\Cf+\alpha$-cluster band on the $0_4^+$ state at $12.9$\ MeV is constructed  
by the developed $\Cf+\alpha$-cluster components in the $\beta>0.8$ region. 

In the negative parity spectra, three dipole states are obtained in the low-energy region;
the $1_1^-$ state with the dominant shell model component, the $1_2^-$ with the $\Cf+\alpha$-cluster
component, and the $1_3^-$ state with the further developed $\Cf+\alpha$-cluster structure. 
The $1_2^-$ and $1_3^-$ states are the band-head states of the lowest and higher-nodal $\Cf+\alpha$-cluster bands, respectively.
They are regarded the parity partners of the positive-parity $\Cf+\alpha$-cluster bands.  
Experimentally, the negative-parity cluster band has not yet been assigned, 
the $1^-$ at 9.19\ MeV and $3^-$ at 9.36\ MeV states observed by $\Cf+\alpha$ scattering experiment~\cite{Avila:2014zwa} are candidate states for the $\Cf+\alpha$-cluster band.

\begin{table}[htbp]
\begin{center}
\begin{tabular}{cccccccc} \toprule
&&Calculation&&&&Experiment& \\
Band&$J^{\pi}_{\textrm{init}}$&$J^{\pi}_{\textrm{fin}}$&$B(E2)$& &$J^{\pi}_{\textrm{init}}$&$J^{\pi}_{\textrm{fin}}$&$B(E2)$ \\ \midrule
Ground band&$2_1^+~(1.30)$&$0_1^+~(0.00)$&1.02&&$2^+~(1.98)$&$0^+~(0.00)$&9.3 \\
&$4_1^+~(2.13)$&$2_1^+~(1.30)$&1.00&&$4^+~(3.56)$&$2^+~(1.98)$&3.3 \\
$\Cf+\alpha$ band&$2_3^+~(9.08)$&$0_2^+~(7.40)$&78.6&&$2^+~(5.26)$&$0^+~(3.64)$&$70\pm42$ \\
 &$4_2^+~(10.6)$&$2_3^+~(9.08)$&87.2&&$4^+~(7.12)$&$2^+~(5.26)$&$15.7\pm4.5$ \\
 &$6_1^+~(14.8)$&$4_2^+~(10.6)$&122&& \\
$\Cf+\alpha$ band &$2_4^+~(13.4)$&$0_4^+~(12.9)$&402&&& \\ 
 (higher nodal) &$4_4^+~(15.0)$&$2_4^+~(13.4)$&718&&& \\ 
 &$6_2^+~(17.5)$&$4_4^+~(15.0)$&818&&& \\ 
\bottomrule
\end{tabular}
\end{center}
\caption{
Calculated and experimental $E2$ transition strengths for the in-band transitions of 
the ground and positive-parity cluster bands.
The experimental data are taken from Ref.~\cite{Tilley:1995zz}.
$B(E2)$ values are shown in units of $e^2\textrm{fm}^4$.
}
\label{tab:E2_positive}
\end{table} 

The calculated $B(E2)$ values for the in-band transitions of the positive- and negative-parity bands 
are listed in Table~\ref{tab:E2_positive} and \ref{tab:E2_negative}, respectively. 
For comparison, the experimental values for the ground and excited ($\Kp=0^+$) bands in positive parity, and those for the strong $E2$ transitions $B(E2)>10$ $e^2$fm$^4$ in negative parity are also listed in the tables.

The calculation underestimates the experimental $E2$ transition strengths in the ground band,
indicating that the present model is insufficient to describe the proton excitations in the $\Oe(2^+_1)$ and 
 $\Oe(4^+_1)$ states.
This result is similar to other AMD calculations~\cite{Furutachi:2007vz,Kanada-Enyo:2019uvg}. 
For the $0_2^+$ band, strong $E2$ transitions are obtained because of the $\Cf+\alpha$ structure. 
The calculated $B(E2;2^+_3\to 0^+_2)$ value is in good agreement with the experimental 
value for the $\Kp=0_2^+$ band. This result supports the assignment of the lowest $\Cf+\alpha$-cluster band
to the experimental $0^+$(3.64~MeV) and $2^+$(5.26~MeV) states.
Further remarkable $E2$ transition strengths are predicted for the higher-nodal $\Cf+\alpha$-cluster band.
In addition, for the negative parity, strong $E2$ transition strengths are obtained in the $\Cf+\alpha$-cluster bands. 
Note that the $E2$ transitions in the lowest $\Cf+\alpha$-cluster band are fragmented into $3^-_{4,5}$ and $5^-_{4,5}$ states. 
Experimentally, significant $E2$ transitions are observed for some states, but the data is insufficient to assign the band structure of cluster states.

\begin{figure}[!h]
\begin{center}
\includegraphics[width=\hsize]{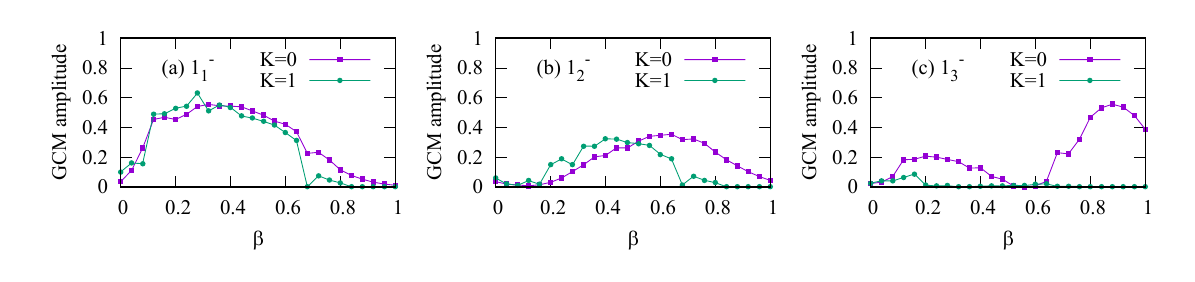}
\end{center}
\caption{GCM amplitudes for the $1^-$ states in $\Oe$.
Panels (a), (b), and (c) show the GCM amplitudes of the $1_1^-$, $1_2^-$, and $1_3^-$ states, respectively.
The $K=0(K=1)$ components for the $\Kp=0^-(\Kp=1^-)$ bases are indicated by squares(circles).
}
\label{fig:GCMamplitude}
\end{figure}

To discuss details of the properties of the $1_1^-$, $1_2^-$, and $1_3^-$ states, 
we show the GCM amplitudes, which are defined by the squared overlap with each base 
in Fig.~\ref{fig:GCMamplitude};
the $K=0$ components of the $\Kp=0^-$ bases and the $K=1$ components of the $\Kp=1^-$ bases 
are presented by squares and circles, respectively.
The shell model bases in $\beta<0.4$ are the dominant component of the $1_1^-$ state.
On the other hand, the $1_2^-$ state has a significant overlap with the $K^\pi=0^-$ component of
the $\Cf+\alpha$ cluster bases; the peak amplitude is observed at $\beta\sim0.6$ (Fig.~\ref{fig:GCMamplitude}~(b)). 
The $1_3^-$ state dominantly contains the $K^\pi=0^-$ $\Cf+\alpha$ cluster components with 
a two-peak structure (Fig.~\ref{fig:GCMamplitude}~(c)), which corresponds to 
the higher-nodal behavior of the $\Cf+\alpha$ cluster mode. 
Note that the $\Kp=1^-$ components in the deformed region ($\beta=$0.4--0.6) 
significantly contribute to the $1_1^-$ and $1_2^-$ states. 
This mixing of the $\Kp=1^-$ components plays an important role in the 
dipole strengths as discussed later.

\begin{table}[htbp]
\begin{center}
\begin{tabular}{cccccccc} \toprule
&&Calculation&&&&Experiment& \\
Band&$J^{\pi}_{\textrm{init}}$&$J^{\pi}_{\textrm{fin}}$&$B(E2)$& &$J^{\pi}_{\textrm{init}}$&$J^{\pi}_{\textrm{fin}}$&$B(E2)$ \\ \midrule
$K^\pi=1^-_1$ &$3_1^-~(6.07)$&$1_1^-~(6.35)$&23.9&&$3^-~(6.40)$&$1^-~(4.46)$& 25$\pm$ 17\\
&$5_1^-~(6.68)$&$3_1^-~(6.07)$&17.8&&$5^-~(8.13)$&$3^-~(5.10)$& 14$\pm$ 14\\
$\Cf+\alpha$ band&$3_4^-~(11.5)$&$1_2^-~(8.68)$&27.4&&$3^-~(8.28)$&$1^-~(4.46)$& 22$\pm$ 22\\
 &$3_5^-~(12.3)$&$1_2^-~(8.68)$&24.7&&&& \\
 &$5_4^-~(13.0)$&$3_4^-~(11.5)$&5.66&&&& \\
 &$5_4^-~(13.0)$&$3_5^-~(12.3)$&10.4&&&& \\
 &$5_5^-~(14.4)$&$3_4^-~(11.5)$&17.1&&&& \\
 &$5_5^-~(14.4)$&$3_5^-~(12.3)$&16.8&&&& \\
$\Cf+\alpha$ band &$3_8^-~(14.4)$&$1_3^-~(8.68)$&105&&& \\ 
 (higher nodal) &$5_6^-~(15.3)$&$3_8^-~(14.4)$&136&&& \\ 
\bottomrule
\end{tabular}
\end{center}
\caption{Calculated and experimental $E2$ transition strengths for the negative parity states.
For the experimental strengths, values of $B(E2)>10\ e^2\textrm{fm}^4$ 
from Ref.~\cite{Tilley:1995zz} are listed.
$B(E2)$ values are shown in units of $e^2\textrm{fm}^4$.
}
\label{tab:E2_negative}
\end{table} 

\begin{figure}[!h]
\begin{center}
\includegraphics[width=\hsize]{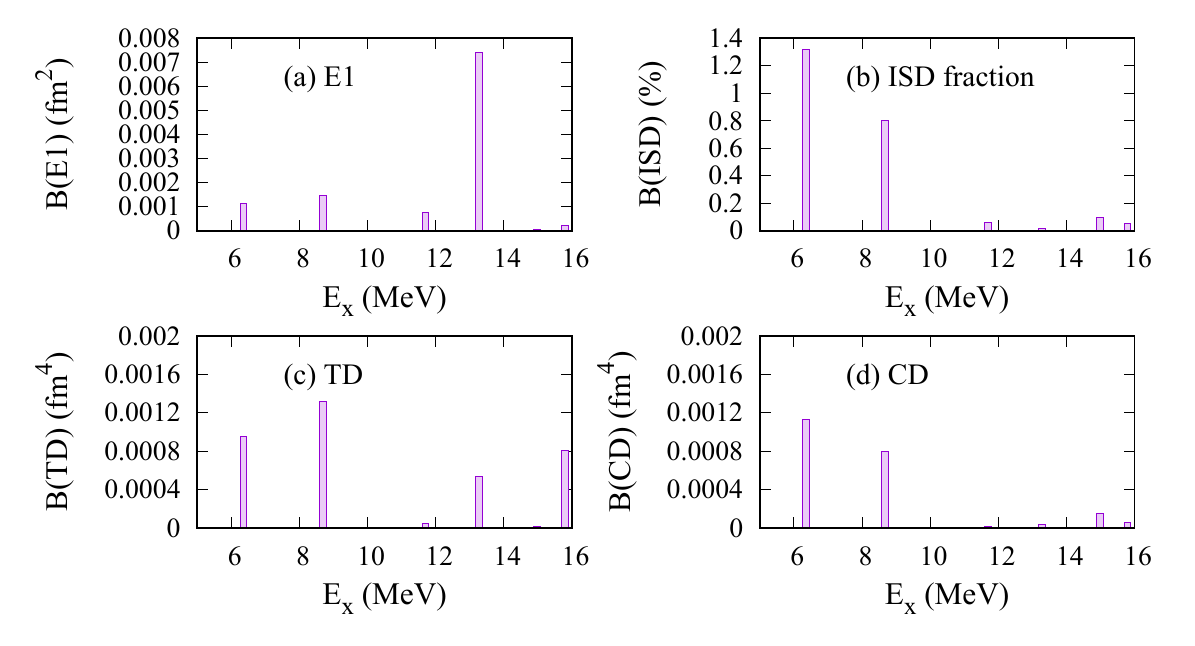}
\end{center}
\caption{
Strength functions of (a) $E1$, (c) TD, and (d) CD transitions for the $0^+_1\to 1^-_k$ transitions. 
The EWSR ratio of the ISD strengths is shown in panel~(b). 
Here, the EWSR values are calculated by using the formula in Ref.~\cite{Harakeh:1981zz}.
}
\label{fig:strength}
\end{figure}

The dipole transition strengths are calculated for the $0^+_1\to 1^-_k$ transitions. 
The strength functions for the $E1$, TD, and CD operators 
are shown in Figs.~\ref{fig:strength}~(a), (c), and (d), respectively. 
The EWSR ratio of the ISD transition strengths
is shown in  Fig.~\ref{fig:strength}~(b).
The three LED states, $1_1^-$, $1_2^-$, and $1_3^-$ have weak $E1$ transition strengths.
The weak $E1$ strength calculated for the $1_1^-$ state is qualitatively consistent with the 
observation, though it is an overestimate compared to the experimental data of 
$B(E1:0_1^+\to1_1^-) < 1.5\times10^{-6}\ \textrm{W.u.}$~\cite{PhysRevC.43.2127} by a few orders of magnitude.
For the TD and CD transitions, the strengths of the $1_1^-$ and $1_2^-$ states are significant, but there is no clear separation of the TD and CD modes in these two LED states.
Unlike these states, the $1_3^-$ state in the higher-nodal cluster band show no remarkable TD or CD strength.

\section{Discussions: Transition current and strength densities for LED}
\label{sec:discussion}
To discuss the detailed properties of the two LED states, the $1_1^-$ and $1_2^-$ states, with significant 
TD and CD transition strengths, we analyze the transition densities in the intrinsic states.
We consider the dominant components of the $0_1^+$, $1_1^-$, and $1_2^-$ states;
the $\Kp=0^+$ base at $\beta=0.28$ labeled $\GSint$ (Fig.~\ref{fig:density}~(a)) for the $0_1^+$ state,
the $\Kp=0^-$ and $\Kp=1^-$ bases at $\beta=0.28$ labeled $\NKzero$ and $\NKone$ (i.e., the normal deformation shown in Fig.~\ref{fig:density}~(d) and (g)), respectively,  for the $1_1^-$ state, 
and the $\Kp=0^-$ base at $\beta=0.64$ labeled $\CKzero$ (the cluster state shown in Fig.~\ref{fig:density}~(e)) for the $1_2^-$ state.
These dominant bases are approximately regarded as the intrinsic states.
Moreover, we analyze the $\Kp=1^-$ base at $\beta=0.52$ labeled $\CKone$ (cluster-like state shown in Fig.~\ref{fig:density}~(h)) because it is contained within the $1_1^-$ and $1_2^-$ states and makes considerable contribution to the dipole strengths.

\begin{figure}[!h]
\begin{center}
\includegraphics[width=12cm]{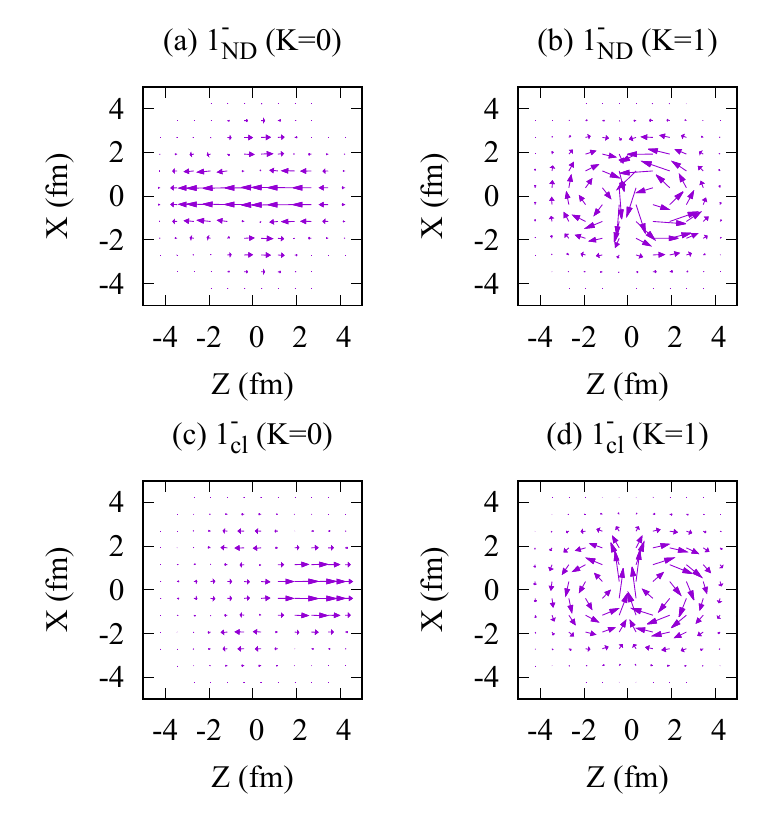}
\end{center}
\caption{
Transition current densities calculated for the intrinsic states.
Panels (a), (b), (c), and (d) show the transition current densities for the $\NKzero$, $\NKone$, $\CKzero$, and $\CKone$ states, respectively.
The vector plot of $\delta\vector{j}_{K}$ on the $Z-X$ plane at $Y=0$ are shown. 
The size of vector is multiplied by 50.
}
\label{fig:current}
\end{figure}

For each intrinsic state, we calculate the transition current density $\delta\vector{j}^K(\vector{r})$
and the local matrix elements ${\cal M}_\textrm{TD,CD}^K(r)$ of the TD and CD operators for the 
excitation from the $\GSint$ state to the $\NKzero$, $\NKone$, $\CKzero$, and $\CKone$ states, which are 
defined as, 
\begin{eqnarray}
\delta\vector{j}^K(\vector{r}) &\equiv & \langle f_K|\vector{j}_{\textrm{nucl}}(\vector{r})|i\rangle ,\\
\hana{M}_{\textrm{TD}}^{K=0}(\vector{r})&=& \frac{1}{c}\left[ (2X^2+2Y^2+Z^2)j^{K=0}_Z - ZXj^{K=0}_X - YZj^{K=0}_Y \right],\label{eq:SD_TD_0} \\ 
\hana{M}_{\textrm{TD}}^{K=1}(\vector{r})&=& \frac{1}{c}\left[ (X^2+2Y^2+2Z^2)j^{K=1}_X - XY j^{K=1}_Y - ZXj^{K=1}_Z \right],\label{eq:SD_TD_1} \\
\hana{M}_{\textrm{CD}}^{K=0}(\vector{r})&=& \frac{1}{c}\left[ -(X^2+Y^2+3Z^2)j^{K=0}_Z - 2ZXj^{K=0}_X - YZj^{K=0}_Y \right],\label{eq:SD_CD_0} \\
\hana{M}_{\textrm{CD}}^{K=1}(\vector{r})&=& \frac{1}{c}\left[ -(3X^2+Y^2+Z^2)j^{K=1}_X - 2XY j^{K=1}_Y - 2ZXj^{K=1}_Z \right],\label{eq:SD_CD_1}
\end{eqnarray}
where $\vector{j}_{\textrm{nucl}}(\vector{r})$ is the convection nuclear current defined in eq.~\eqref{eq:current}.
The initial state is projected onto $K=0$ as $|i\rangle= \hat{P}_{K=0}|\GSint\rangle$, and
the final states $|f_K \rangle$ are the $K$-projected intrinsic states, 
$\hat{P}_{K=0}|\NKzero\rangle$, $\hat{P}_{K=1}|\NKone\rangle$, $\hat{P}_{K=0}|\CKzero\rangle$, and $\hat{P}_{K=1}|\CKone\rangle$.
Note that $\hana{M}_{\textrm{TD}}^{K}(\vector{r})$ and $\hana{M}_{\textrm{CD}}^{K}(\vector{r})$
correspond to the integrand of the TD and CD transition matrix elements and are termed TD and CD strength density, respectively, in this paper. 

\begin{figure}[!h]
\begin{center}
\includegraphics[width=12cm]{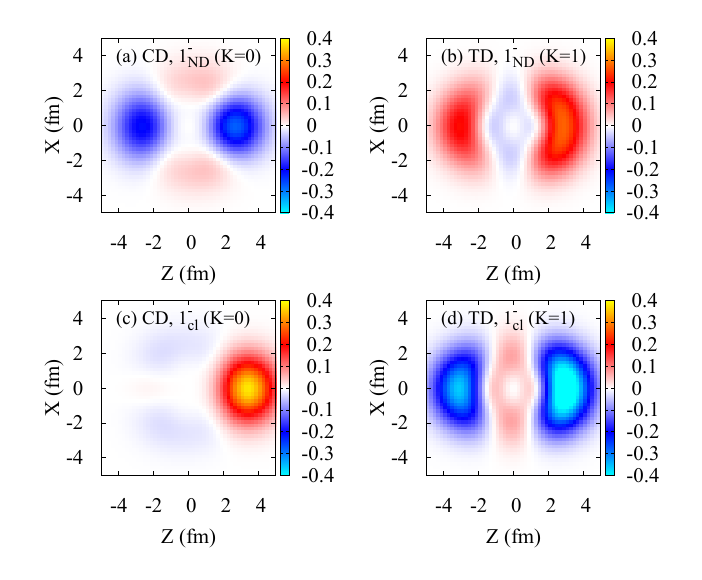}
\end{center}
\caption{
TD and CD strength densities calculated for the intrinsic states.
Panels~(a) and (c) show the CD strength densities for the $\NKzero$ and $\CKzero$ states, respectively, and panels~(b) and (d) present the TD strength densities for the $\NKone$ and $\CKone$ states, respectively.
These heat maps for the strength densities on the $Z-X$ plane at $Y=0$ are plotted.
}
\label{fig:strength_distribution}
\end{figure}

The transition current densities calculated for the $\NKzero$, $\NKone$, $\CKzero$, and $\CKone$ states are shown in Fig.~\ref{fig:current}. 
The TD and CD strength densities are shown in Fig.~\ref{fig:strength_distribution}.

In the transitions to the $\NKzero$ and $\NKone$ bases for the $1_1^-$ state, 
the 1p-1h excitations generate the transition current
in the internal region, namely, the translational current in the $Z$-direction in the $\NKzero$ base  (Fig.~\ref{fig:current}~(a)) and the vortical current mainly from the proton part 
in the $\NKone$ base (Fig.~\ref{fig:current}~(b)). 
These currents in the $\NKzero$ and $\NKone$ bases give the significant CD and TD strength densities as
shown in Figs.~\ref{fig:strength_distribution}~(a) and \ref{fig:strength_distribution}~(b) contributing to the CD and TD strengths for the $1_1^-$ state, respectively. 
In other words, the origin of the CD and TD strengths in the $0^+_1\to 1_1^-$ excitation are the $K=0$ and $K=1$ 1p-1h modes, respectively, on the top of the normal deformation.

For the $1_2^-$ state, we first discuss the transition properties of its main component, i.e., the $\CKzero$ base.
In the transition to this $\CKzero$ base, a remarkable translational current is 
produced in the outer region around $Z\sim2.5$\ fm by the motion of the $\alpha$-cluster (Fig.~\ref{fig:current}~(c)).
This translational motion results in remarkable CD strength densities as shown in Fig.~\ref{fig:strength_distribution}~(c) and contributes to the significant CD strength in the $1_2^-$ excitation. 
Next, we discuss the properties of the $\CKone$ component, which is significantly mixed in the 
the $1_2^-$ state.
As can be seen in 
Figs.~\ref{fig:current}~(d) and \ref{fig:strength_distribution}~(d)
for this $\CKone$ base, the remarkable vortical current and TD strength density are 
obtained,  in particular, in the surface region.  
This vortical current of the $\CKone$ component is a major origin of the TD strength in the $1_2^-$ state.
It is worth to mention that this $\CKone$ component is also mixed in the $1_1^-$ state and 
enhances the TD strength of the $1_1^-$ state. 

From these analyses, the CD and TD transition properties of the two LED states can be roughly understood by
two kinds of excitation modes: the 1p-1h excitation in the $1_1^-$ state and the cluster excitation in the 
$1_2^-$ state. In both modes of the 1p-1h and cluster excitations, the CD and TD 
transition strengths are generated in the $K^\pi=0^-$ and $K^\pi=1^-$ components 
of the deformed bases, respectively. In particular, remarkably strong surface currents 
are generated by the $\alpha$ cluster motion in the cluster excitation. 
This collective~(cluster) excitation in the largely deformed bases 
further enhances the CD and TD transition strengths compared with the 1p-1h excitations in the 
normal deformation.
This plays an important role in the properties of the LED states.
As shown in Figs.~\ref{fig:GCMamplitude}~(a) and (b), for the GCM amplitudes, 
the cluster bases around $\beta=0.5$--$0.6$ are significantly mixed in the $1^-_1$ and $1^-_2$ states. 
Through the configuration mixing along $\beta$, these cluster components significantly 
enhance the CD and TD strengths in the two LED states.
Moreover, as a result of the $K$ and configuration mixing, the CD and TD natures do not separately
appear in the independent $1^-$ states of $\Oe$. 
This is different from the decoupling case obtained for $\Beten$ with a large deformation~\cite{Shikata:2020lgo}.
Similar mode mixing features were obtained for $\Os$ in our previous study~\cite{Shikata:2020lgo}.
Such features are expected in other neutron-rich oxygen isotopes too.

\section{Summary}\label{sec:summary}

We investigated the LED excitations in $\Oe$ using $\beta$-AMD with K-VAP method.
In the AMD result, the shell model, $\Cf+\alpha$, and higher nodal $\Cf+\alpha$ bases are obtained in $\Kp=0^{\pm}$ bases.
Through GCM, two LED states, the $1_1^-$ and $1_2^-$ states, are obtained.
The main components of the former state are  shell model bases with small deformation and those of the 
latter state are the $\Cf+\alpha$ cluster bases with large deformation.
Moreover, the $1_3^-$ is obtained as the higher nodal $\Cf+\alpha$ cluster state.
For the $1^-_1$ and $1^-_2$ states, the significant TD and CD strengths are obtained,
whereas the strengths are weak for the $1^-_3$. 

Detailed analyses of the transition properties of the $1_1^-$ and $1_2^-$ states 
were performed by calculating the transition current densities and strength densities.
For the $1_1^-$ state, the TD and CD strengths mainly originate from the 1p-1h excitation of the shell model bases.
For the $1_2^-$ state, the significant CD strength is generated by the $\alpha$-cluster motion
in the $\Kp=0^-$ components of the $\Cf+\alpha$ cluster bases, and the TD strengths originate from 
the $\Kp=1^-$ components of the $\Cf+\alpha$ cluster bases through the $K$ and configuration mixing.
The cluster components also contribute to the $1_1^-$ state; they provide additional enhancement of 
the CD and TD strengths of the $1_1^-$ state via the configuration mixing of cluster bases to the 
dominant shell model bases.
Thus, the $\Cf+\alpha$ cluster excitation plays important roles in the transition properties of the two LED states
of $\Oe$.

Nesterenko {\it et al.} investigated the low-lying TD and CD modes in deformed nuclei and 
showed the clear separation of the TD and CD modes in LED states in largely deformed systems.
However,  the present result does not show such a clear separation.
Instead, the two modes mix in the $1^-_1$ and $1^-_2$ states because the $\Oe$ system favors small deformation.
The similar scenario can be extended to the $^{20}$O system. 
Indeed, fragmentation of the CD strengths to the $1^-_1$ and $1^-_2$ states of $^{20}$O  has been reported by  
the recent experiment by Nakatsuka {\it et al.}~\cite{Nakatsuka:2017dhs}.
Application of the present AMD method with K-VAP to $^{20}$O is a challenge for the future to clarify the LED modes in neutron-rich oxygen isotopes.

\section*{Acknowledgment}

The authors thank to Dr.~Nesterenko and Dr.~Chiba for fruitful discussions.
The computational calculations of this work were performed by using the
supercomputer in the Yukawa Institute for theoretical physics, Kyoto University. 
This work was supported by 
JSPS KAKENHI Grant Nos. 18J20926, 18K03617, and 18H05407.

\bibliographystyle{ptephy}
\bibliography{reference-IMANUM} 

\end{document}